\documentclass[twocolumn,showpacs,preprintnumbers,amsmath,amssymb,prl]{revtex4}

\usepackage{graphicx}
\usepackage{bm}

\begin{document}

\title{Breakdown of the Rosenfeld Excess Entropy Scaling Relations for the Core-Softened Systems:
Thermodynamic Path Dependence}

\author{Yu. D. Fomin}
\affiliation{Institute for High Pressure Physics, Russian Academy
of Sciences, Troitsk 142190, Moscow Region, Russia}

\author{V. N. Ryzhov}
\affiliation{Institute for High Pressure Physics, Russian Academy
of Sciences, Troitsk 142190, Moscow Region, Russia}

\date{\today}

\begin{abstract}
We analyze the applicability of the Rosenfeld entropy scaling
relations to the systems with the core-softened potentials
demonstrating the water-like anomalies. It is shown that the
validity of the of Rosenfeld scaling relation for the diffusion
coefficient depends on the thermodynamic path which is used for
the calculations of the kinetic coefficients and the excess
entropy. In particular, it is valid along isochors, but it breaks
down along isotherms.
\end{abstract}

 \maketitle

In 1977 Rosenfeld proposed the relations connecting transport
properties of a liquid with the excess entropy \cite{ros1}. In
order to write down these relations one should use the reduced
forms of the transport coefficients:
\begin{equation}
 D^*=D \frac{\rho ^{1/3}}{(k_BT/m)^{1/2}}
\end{equation}

\begin{equation}
 \eta^*=\eta \frac{\rho^{-2/3}}{(mk_BT)^{1/2}}
\end{equation}
where $D$ and $\eta$ are the diffusion coefficient and the
viscosity. According to the Rosenfeld suggestion, the reduced
transport coefficients can be expressed in the form
\begin{equation}
  X=a_X \cdot e^{b_X S_{ex}},
\end{equation}
where $S_{ex}=(S-S_{id})/{(Nk_B)}$ is excess entropy of the
liquid, $X$ is the transport coefficient, and $a_X$ and $b_X$ are
the constants which depend on the studying property \cite{ros2}.
Interestingly, the coefficients $a$ and $b$ show an extremely weak
dependence on the material and can be considered as universal.

In his original works Rosenfeld considered hard spheres, soft
spheres, Lennard-Jones system and one-component plasma
\cite{ros1,ros2}. After that the excess entropy scaling was
applied to many different systems including core-softened liquids
\cite{errington1,errington2,india1,indiabarb,weros}, liquid metals
\cite{liqmet1,liqmet2}, binary mixtures \cite{binary1,binary2},
ionic liquids \cite{india2,ionicmelts}, network-forming liquids
\cite{india1,india2}, water \cite{buldwater}, chain fluids
\cite{chainfluids} and bounded potentials
\cite{weros,klekelberg,klekelberg1}.

Nevertheless, controversies still remain. For example, up to the
moment it is not clear whether the Rosenfeld scaling relations are
applicable to the core softened systems. Some publications state
that the scaling relations are valid for such systems
\cite{errington2,indiabarb}, while in our recent work it was shown
that the scaling relations may break down for the core softened
systems \cite{weros}. This article presents a discussion of this
contradiction. Basing on the molecular dynamics simulations of the
two core-softened systems we show that the validity of the
Rosenfeld scaling relations depends on the thermodynamic path
which is used for the calculations of the kinetic coefficients and
the excess entropy. In particular, the exponential functional form
of the relation between the excess entropy and the reduced
diffusion coefficient holds along isochors while along isotherms
one observes its breakdown.


Two systems are studied in the present work. The first one is the
core-softened system introduced by de Oliveira et al
\cite{barbosapot}. This system is described by the spherically
symmetric potential represented by a sum of a Lennard-Jones
contribution and a Gaussian-core interaction (LJG):
\begin{equation}
  U(r)=4\varepsilon
  \left[\left(\frac{\sigma}{r}\right)^{12}-\left(\frac{\sigma}{r}\right)^{6}\right]
  +a\varepsilon
  \cdot \exp\left[-\frac{1}{c^2}\left(\frac{r-r_0}{\sigma_0}\right)^2\right],
\end{equation}
with $a=5.0$, $r_0/ \sigma=0.7$ and $c=1.0$. This model can
qualitatively reproduce water's density, diffusivity, and
structural anomalies. The diffusivity of this system was studied,
for example, in the papers \cite{barbosapot,indiabarb}. In the
papers \cite{errington2,indiabarb} the validity of the Rosenfeld
scaling relation for this system was checked. In this paper we
discuss these results and compare them with our calculations.

The second system studied in this work is a repulsive shoulder
system (RSS) introduced in our previous work \cite{wejcp}. This
system has a potential
\begin{equation}
  U(r)=
  \left(\frac{\sigma}{r}\right)^{14}+\frac{1}{2}\varepsilon
  \cdot[1-\tanh(k_0\{r-\sigma_1\})],
\end{equation}
where $\sigma$ is the "hard"-core diameter, $\sigma_1=1.35$ is the
soft-core diameter, and $k_0=10.0$. In Ref. \cite{wepre} it was
shown that this system demonstrates anomalous thermodynamic
behavior. In our previous publication \cite{weros} the Rosenfeld
relation for this system was studied. It was shown that the
scaling relation for the diffusion coefficient breaks down for
this system in the anomalous diffusion region.


In this paper we use the dimensionless quantities: $\tilde{{\bf
r}}={\bf r}/ \sigma$, $\tilde{P}=P \sigma
^{3}/\varepsilon ,$ $\tilde{V}=V/N \sigma^{3}=1/\tilde{\rho},$ $\tilde{T}%
=k_{B}T/\varepsilon $. Since we use only these reduced units we
omit the tilde marks.

For the investigation of the LJG potential we simulate a system of
$1000$ particles in a cubic box for the densities ranging from
$\rho=0.01$ till $\rho=0.35$ with the step $\delta \rho =0.01$.
The time step used is $dt=0.001$. The equilibration period
consists of $1 \cdot 10^6$ time steps and the production period -
$2.5 \cdot 10^6$ time steps. During the equilibration the
temperature is kept constant by velocity rescaling while during
the production cycle $NVE$-MD is used. The equations of motion are
integrated by velocity-Verlet algorithm. The following isotherms
are simulated: $T=0.2;0.3;0.4;0.5;0.6;1.0$ and $1.5$.

The simulation setup of the RSS was described in Ref.
\cite{weros}.

The excess entropy in both cases was computed via thermodynamic
integration method. For doing this we calculate excess free energy
by integrating the equation of state along an isotherm:
$\frac{F_{ex}}{Nk_BT}=\frac{F-F_{id}}{Nk_BT}=\frac{1}{k_BT}
\int_0^{\rho} \frac{P(\rho ')-\rho ' k_BT}{\rho '^2} d\rho'$. The
excess entropy is computed via $S_{ex}=\frac{U-F_{ex}}{N k_BT}$.


\begin{figure}
\includegraphics[width=7cm, height=7cm]{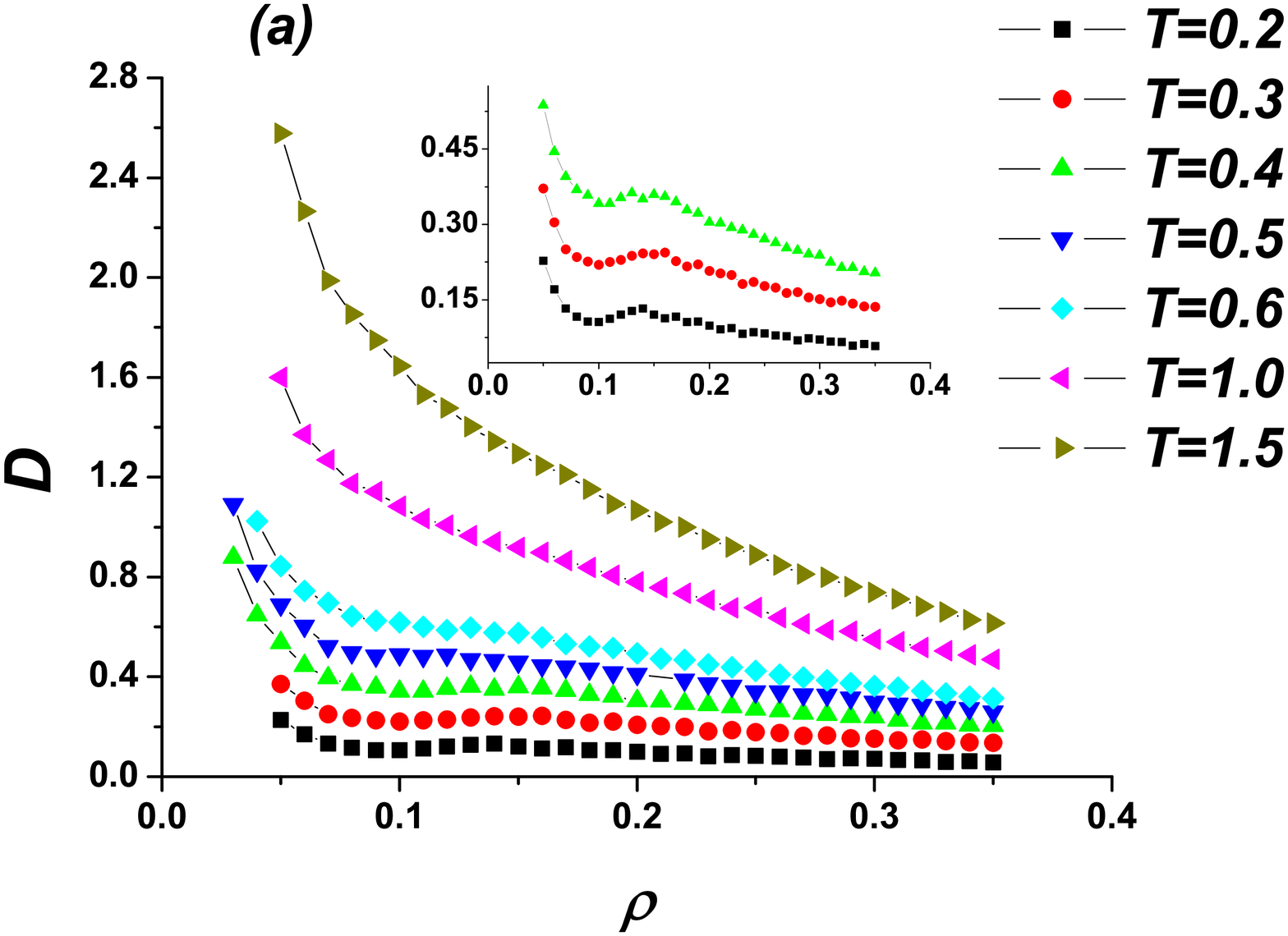}%

\includegraphics[width=7cm, height=7cm]{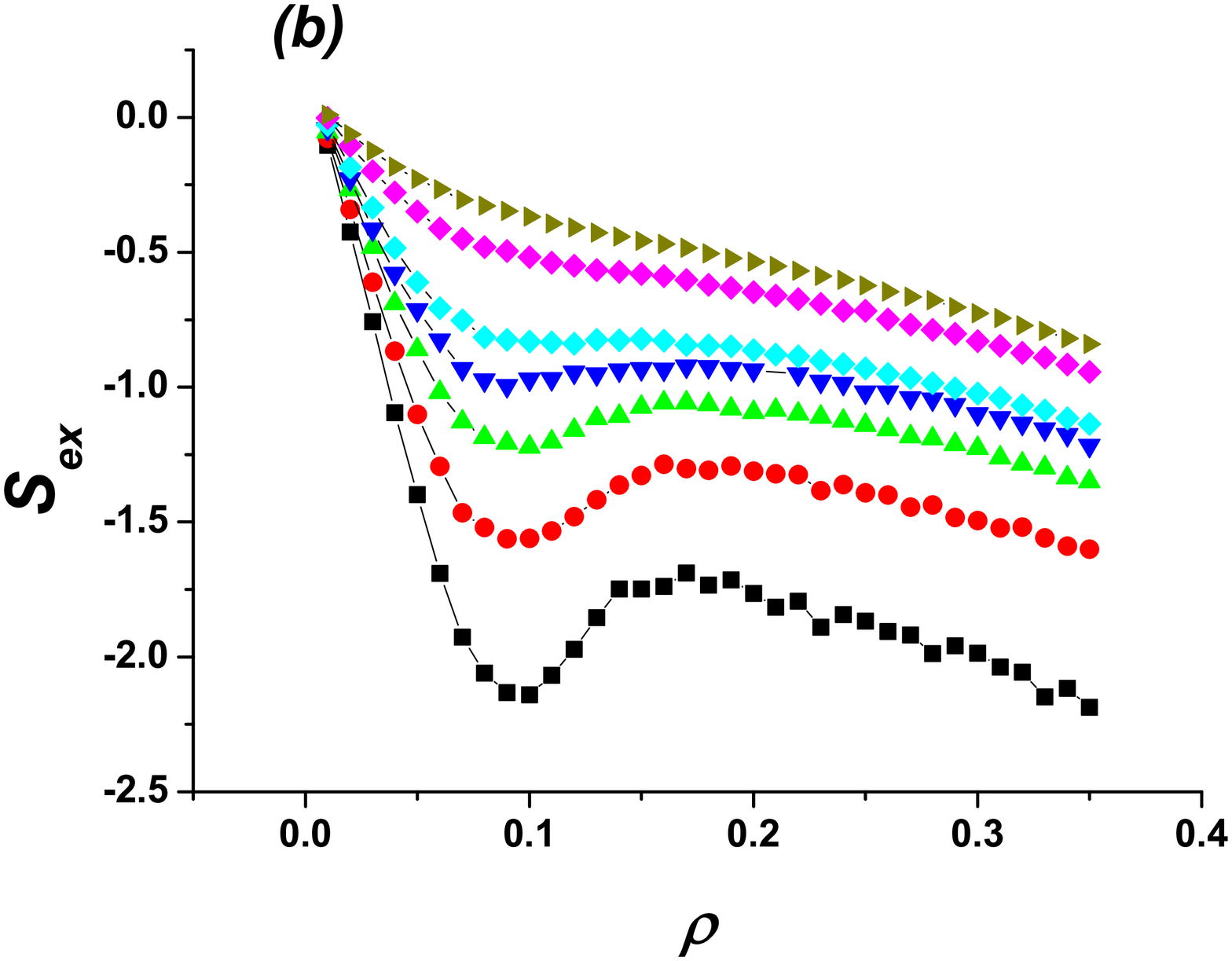}%
\caption{\label{fig:fig1} (a) The diffusion coefficient of the LJG
system along some isotherms. (b) The excess entropy of the LJG
system for a set of isotherms. (Color online)}
\end{figure}

\begin{figure}
\includegraphics[width=6cm, height=6cm]{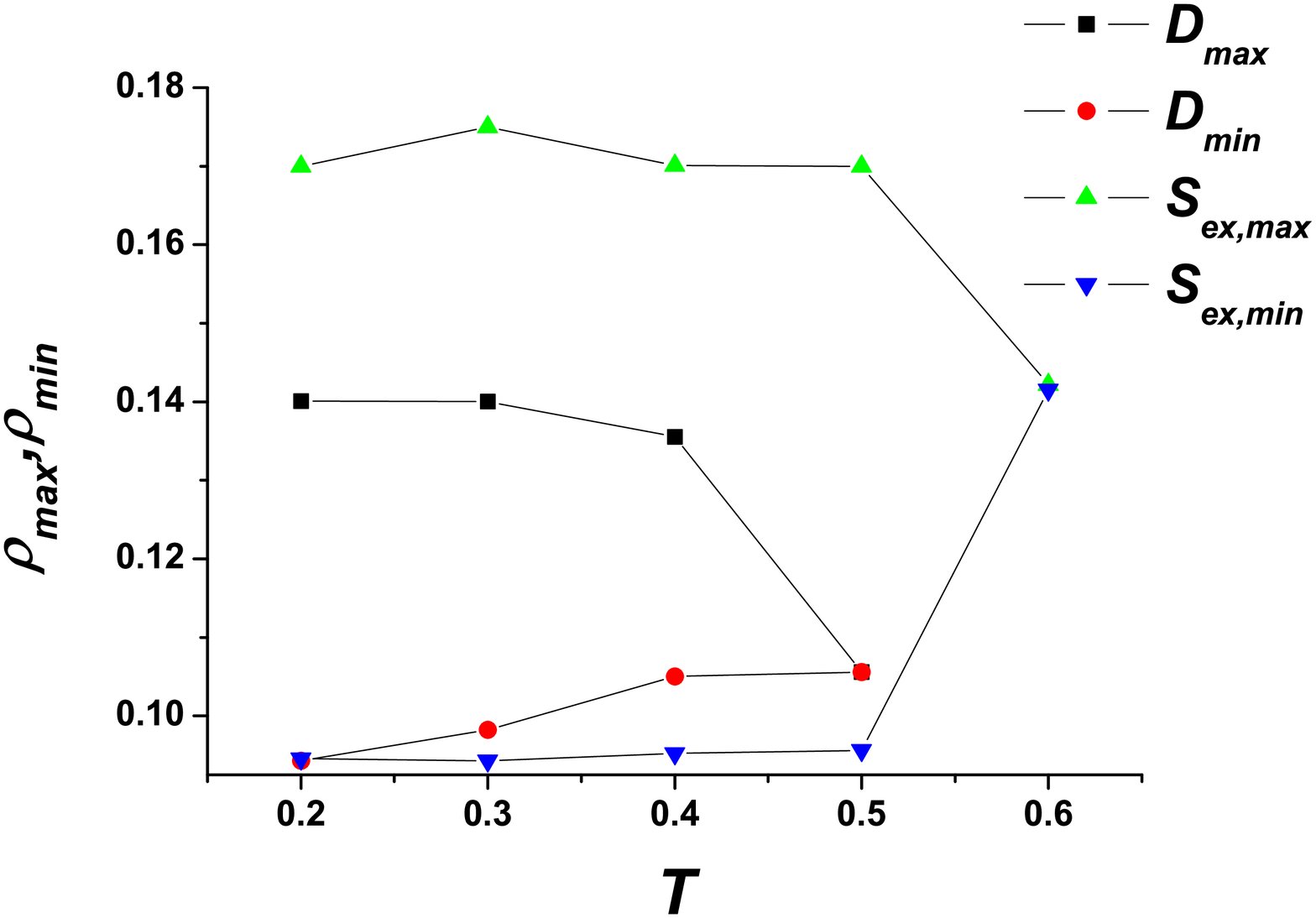}%

\caption{\label{fig:fig2}  The densities of maximum and minimum of
the diffusivity and the excess entropy for the several isotherms.}
\end{figure}

Fig.~\ref{fig:fig1}(a) presents the diffusion coefficient of the
LJG system along a set of isotherms. One can see that at low
temperatures the diffusion coefficient shows nonmonotonic behavior
which is known as the diffusion anomaly. Fig.~\ref{fig:fig1}(b)
demonstrates the excess entropy for the same set of isotherms.
From these figures it follows that the diffusion coefficient and
the excess entropy along an isotherm have similar qualitative
behavior. However, the location of the extremum points of
diffusivity and excess entropy is different (Fig.~\ref{fig:fig2}).
It means that there are some regions where one function increases
while another one decreases and vice versa. Clearly, this kind of
behavior can not be consistent with the Rosenfeld scaling formula.
From this one can conclude that the Rosenfeld scaling relation is
not applicable to the diffusivity along an isotherm for the LJG
model.

\begin{figure}
\includegraphics[width=7cm, height=7cm]{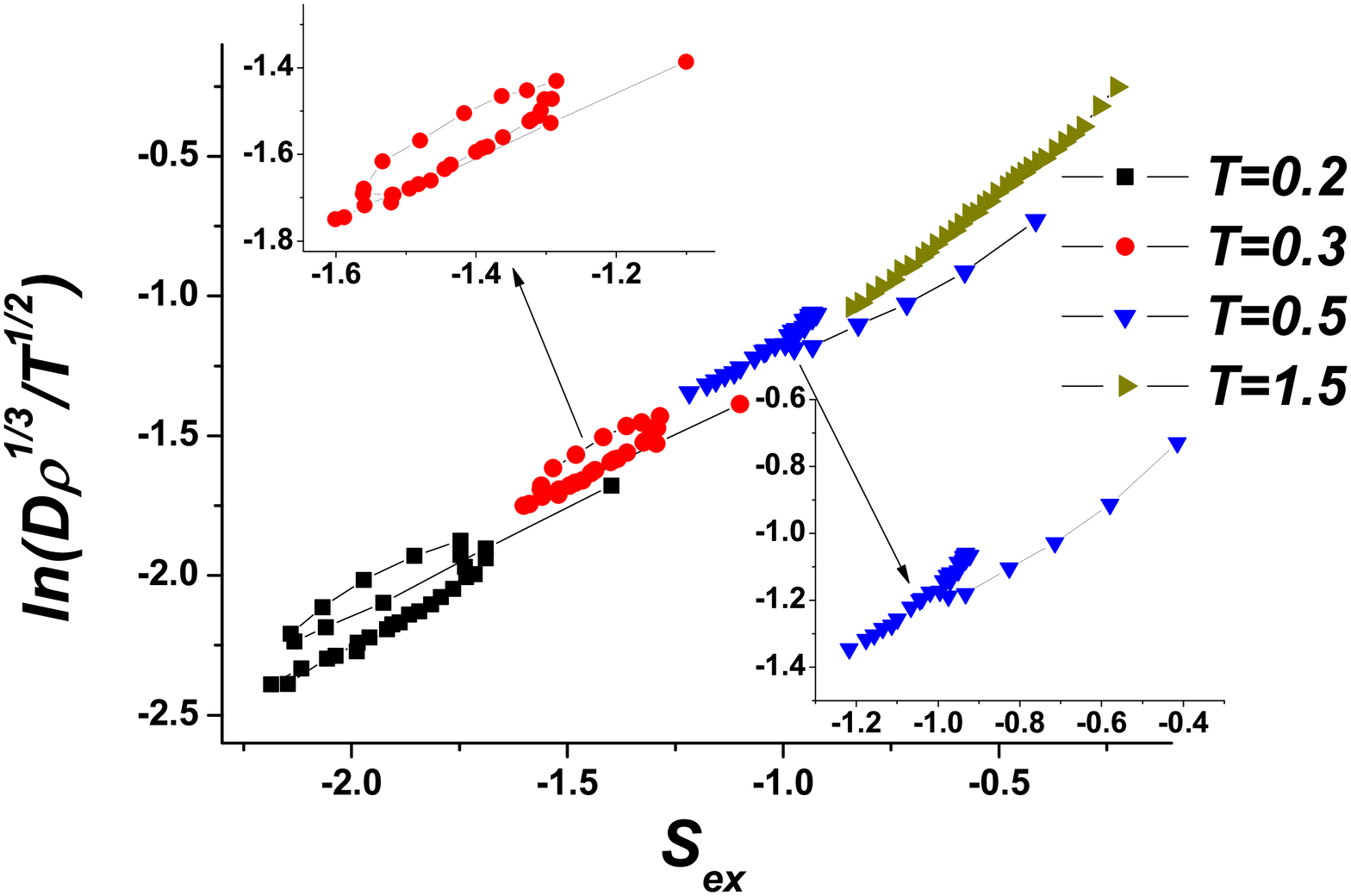}%

\caption{\label{fig:fig3} The logarithm of the reduced diffusion
coefficient for the LJG model along the several isotherms. The
insets correspond to the temperatures $T=0.3$ (upper panel) and
$T=0.5$ (lower panel) in the larger scale. (Color online)}
\end{figure}

Fig.~\ref{fig:fig3} presents the logarithm of reduced diffusion
coefficient vs excess entropy along a set of isotherms. One can
see that at low temperatures the curves demonstrate a selfcrossing
loop like the one observed for the RSS model in Ref. \cite{weros}.
This loop becomes less pronounced with increasing the temperature
and at $T=0.5$ the self crossing disappears. At high temperatures
the curve comes to the straight line limit which corresponds to
the small influence of the soft core at such temperatures.

\begin{figure}
\includegraphics[width=7cm, height=7cm]{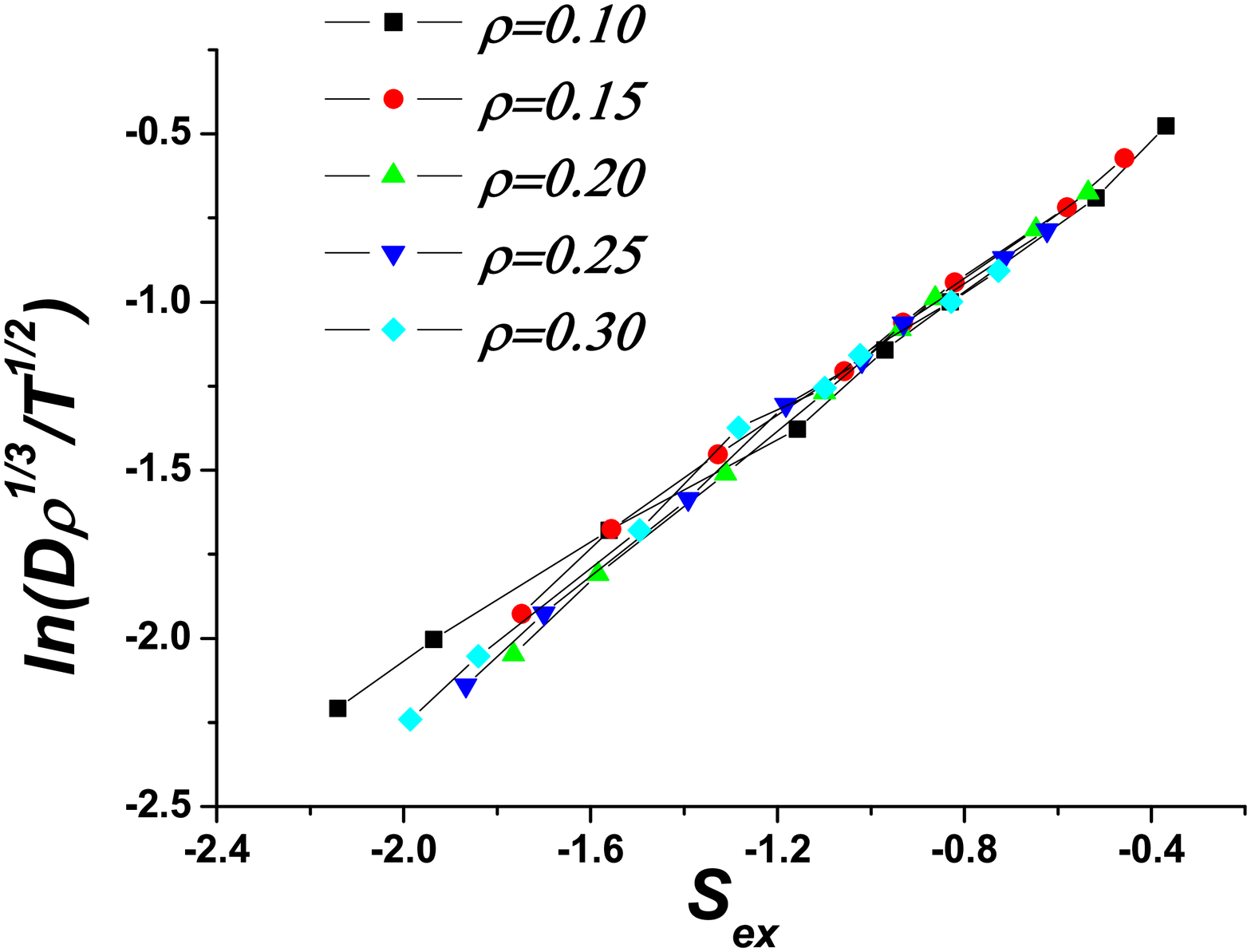}%

\caption{\label{fig:fig4} The logarithm of the reduced diffusion
coefficient for the LJG model along the several isochors. (Color
online)}
\end{figure}

It can be seen from Figs.~\ref{fig:fig1}(a) and (b), that both the
diffusivity and the excess entropy along isochors are monotonous
(this corresponds to the vertical lines through the data points).
It allows to expect that the exponential relation between the
diffusion coefficient and the excess entropy holds along isochors.
Fig.~\ref{fig:fig4} shows the $\ln(D^*)$ vs $S_{ex}$ along a set
of isochors. From this figure one can see that all curves with
good accuracy correspond to the Rosenfeld scaling relation. This
result is in agreement with with Refs.
\cite{errington2,indiabarb}.

\begin{figure}
\includegraphics[width=7cm, height=7cm]{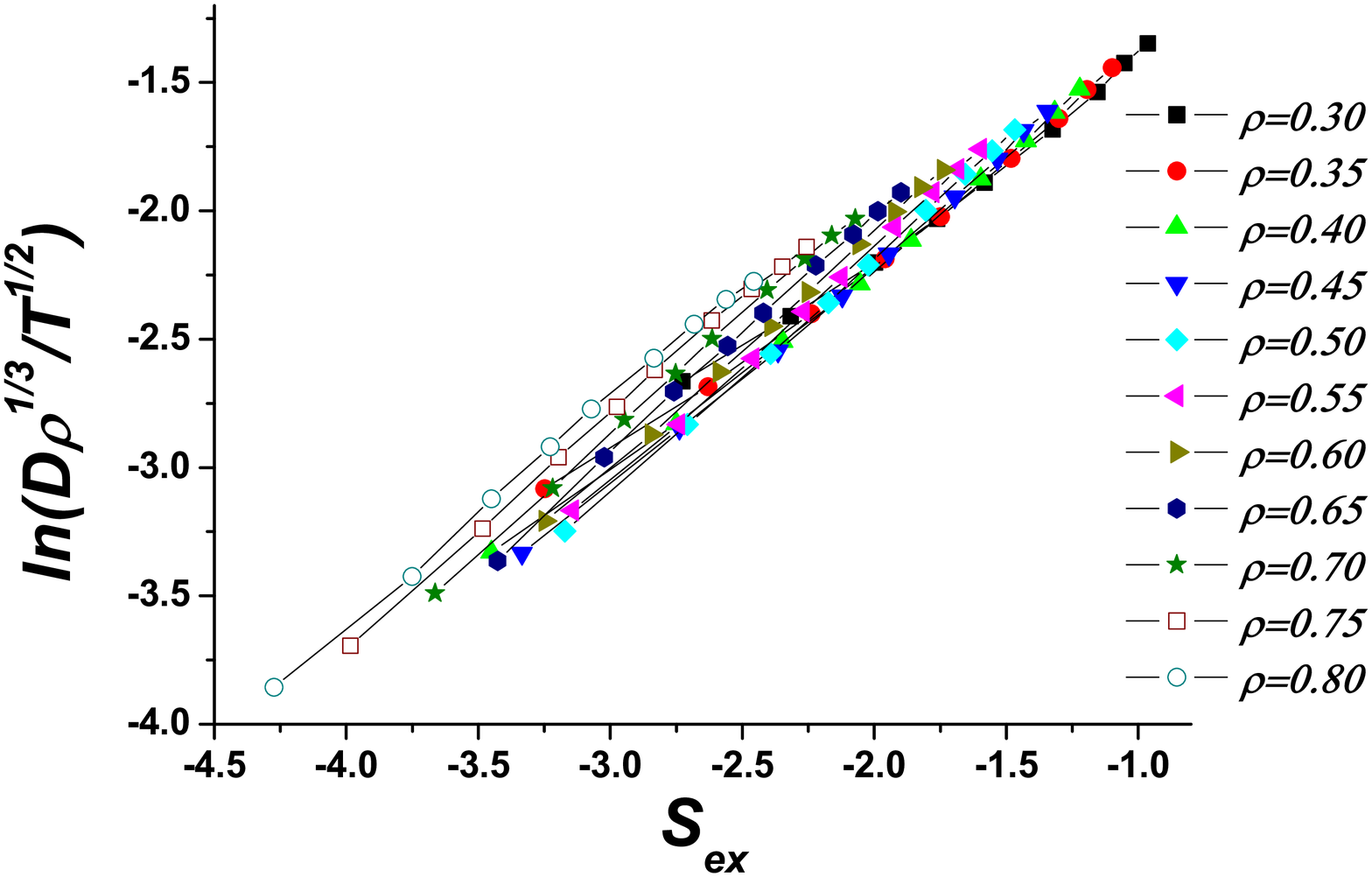}%

\caption{\label{fig:fig5} Reduced diffusion logarithm for RSS
along a set of isochors.}
\end{figure}

In order to show that the dependence of the Rosenfeld scaling
relations on the thermodynamic path is a general property of the
core-softened systems we study another core softened potential
introduced above - the repulsive shoulder system (RSS). In our
previous publications we showed that this system demonstrates
anomalous diffusion behavior at low temperatures \cite{wepre} and
that the Rosenfeld scaling relation along isotherms is not
applicable \cite{weros}: the curves demonstrate the self crossing
loops like in the case of the LJG system. Fig.~\ref{fig:fig5}
shows the logarithm of the reduced diffusion coefficient along a
set of isochors for the RSS. As it can be seen from this figure,
the dependence of $D^*$ on $S_{ex}$ is linear. However, the slope
of the line shows an isochor dependence. Fig.~\ref{fig:fig5} shows
that the slope remains approximately constant for low densities
($\rho=0.3-0.55$) while on increasing the density the slope also
increases.

\begin{figure}
\includegraphics[width=7cm, height=7cm]{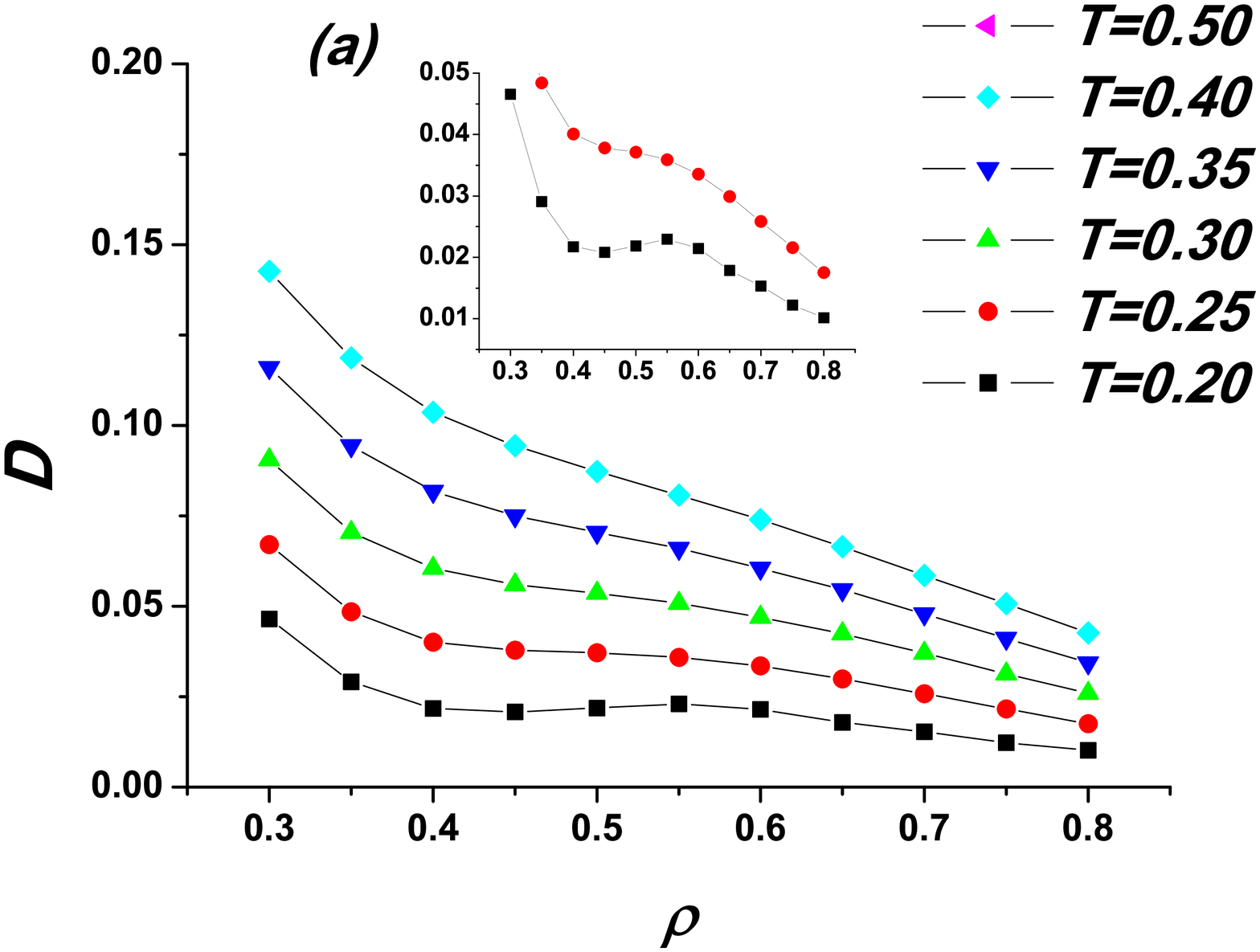}%

\includegraphics[width=7cm, height=7cm]{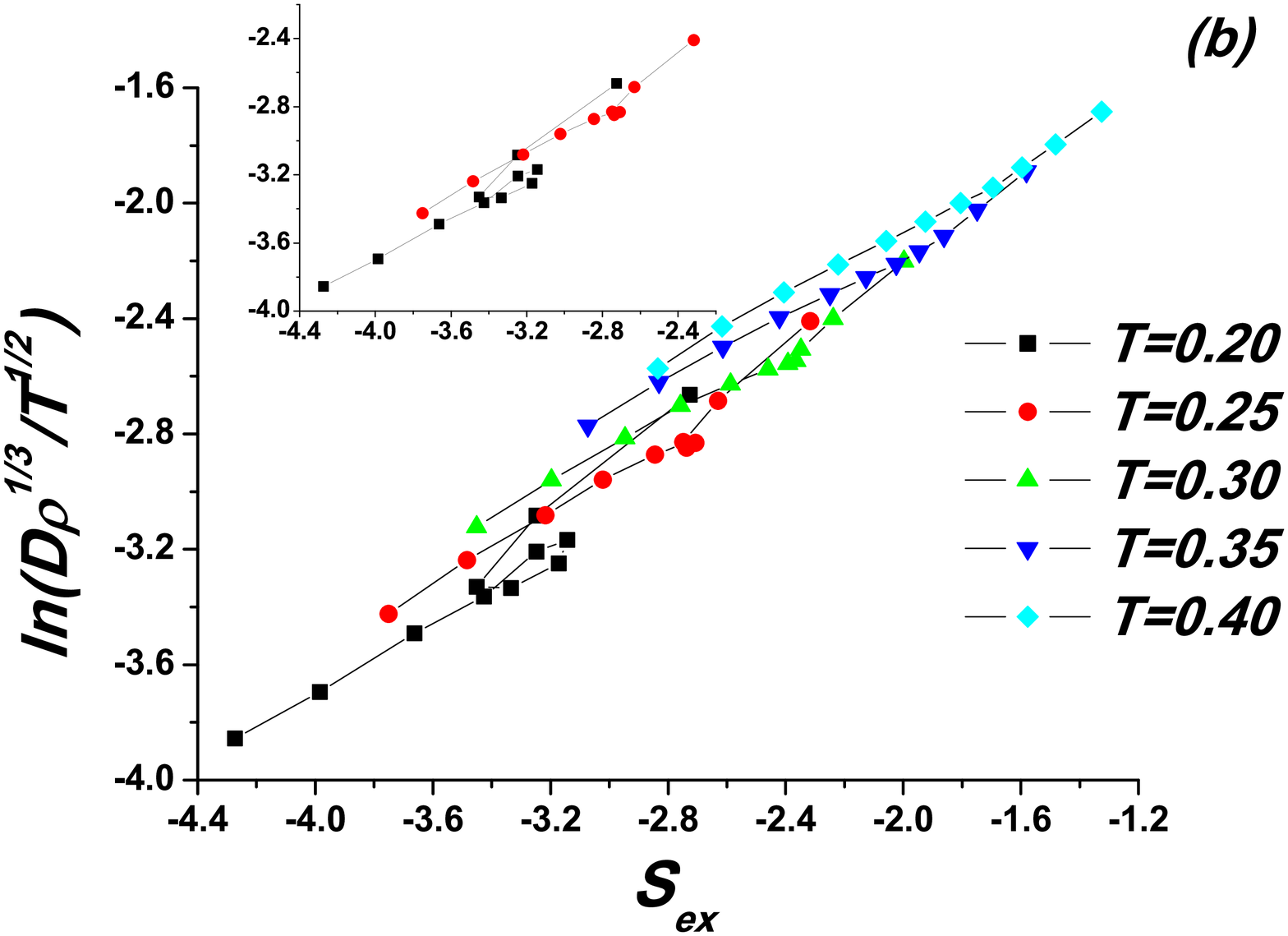}%
\caption{\label{fig:fig6} (a) The diffusion coefficient for the
RSS system along several isotherms. The inset enlarges the
isotherms $T=0.2$ (squares) and $T=0.25$ (circles). (b) The
logarithm of the reduced diffusion coefficient as a function of
$S_{ex}$ along a set of isotherms for RSS. (Color online).}
\end{figure}

The diffusion coefficient along isotherms as a function of the
density and the reduced diffusion coefficient as a function of
excess entropy are shown in Figs.~\ref{fig:fig6}(a) and (b). As it
can be seen from Fig.~\ref{fig:fig6}(a), the density $\rho=0.55$
corresponds to the density maximum at $T=0.2$. In Ref.
\cite{wejcp} it was shown that this system is effectively
quasibinary. The reason for the complex behavior of this system is
related to the competition between two length scales $\sigma$ and
$\sigma_1$. This allows to suggest that the density $\rho=0.55$
belongs to the boundary between two different regimes. It seems
that the densities from $\rho=0.6$ till $\rho=0.8$ belong to a
cross section region between low- and high-density regimes which
leads to the shift of all curves with respect to each other. One
can expect that at higher densities the logarithm of $D^*$ comes
to a straight line again but the slope of this line should be
different from that at low densities.

In conclusion, in the present article we carry out a study of
applicability of the Rosenfeld excess entropy scaling relation for
the diffusion coefficient in the case of the systems with the
core-softened potentials. We find that the validity of the of
Rosenfeld scaling relation depends on the thermodynamic path. In
particular, it is valid along isochors (although the scaling
parameters show the isochor dependence), but it breaks down along
isotherms. Interestingly, both the excess entropy and the
diffusivity are monotonous along isochors while along isotherms
they show the anomalous increase at low temperatures. It is this
inconsistency in the location of the minimum and maximum points of
the diffusion coefficient and the excess entropy along the
isotherms which leads to the breakdown of Rosenfeld scaling
relation for these systems. This allows to suggest that the
scaling relation is valid for the core-softened systems along any
thermodynamic path if both $S_{ex}$ and $D^*$ are monotonous along
it. However, this suggestion requires further investigation.

Another class of systems which attract a lot of attention with
respect to the Rosenfeld scaling relation is bounded potential
systems \cite{weros,klekelberg,klekelberg1}. It is important to
note that the conclusion of this paper is not applicable to these
systems. For example, in Refs. \cite{klekelberg,klekelberg1}
scaling relation for the Gauss Core Model (\cite{gaussstill})
along isochors was studied. According to these publications, the
curves deviate from exponential dependence. This allows to suggest
that the relation between the dynamic and thermodynamic properties
for the bounded potential systems is different from the systems
with the hard core.


\begin{acknowledgments}
We thank V. V. Brazhkin and Daan Frenkel for stimulating
discussions. Our special thanks to Prof. Ch. Chakravarty who
attracted our attention to the problems considered here. Y.F. also
thanks the Russian Scientific Center Kurchatov Institute for
computational facilities. The work was supported in part by the
Russian Foundation for Basic Research (Grants No 08-02-00781 and
No 10-02-00700) and Russian Federal Program 02.740.11.5160.
\end{acknowledgments}


\begin{thebibliography}{99}


\bibitem{ros1} Ya. Rosenfeld, Phys. Rev. A, \textbf{15}, 2545
(1977).

\bibitem{ros2} Ya. Rosenfeld, J. Phys.: Condens. Matter
\textbf{11}, 5415 (1999).

\bibitem{india1} R. Sharma, S. N. Chakraborty, and Ch. Chakravartya J. Chem. Phys. \textbf{125}, 204501
(2006).

\bibitem{errington1} J. R. Errington, Th. M. Truskett, J. Mittal,
J. Chem. Phys. \textbf{125}, 244502 (2006)

\bibitem{errington2} J. Mittal, J. R. Errington, Th. M. Truskett,
J. Chem. Phys. \textbf{125}, 076102 (2006).

\bibitem{indiabarb} A. B. de Oliveira, E. A. Salcedo Torres, Ch. Chakravarty, M. C. Barbosa,
arXiv:1002.3781 (2010).


\bibitem{weros} Yu. D. Fomin, N. V. Gribova, V. N. Ryzhov, arXiv:
1001.0111 (2009).

\bibitem{liqmet1} A. Samanta, Sk. Musharaf Ali, S. K. Ghosh, J.
Chem. Phys. \textbf{123}, 084505 (2005).

\bibitem{liqmet2} J. J. Hoyt, Mark Asta, and Babak Sadigh, Phys.
Rev. Lett. \textbf{85}, 594 (2000).

\bibitem{binary1} A. Samanta, Sk. Musharaf Ali, and S. K. Ghosh,
Phys. Rev. Lett. \textbf{87}, 245901 (2001).

\bibitem{binary2} A. Samanta, Sk. Musharaf Ali, and S. K. Ghosh,
Phys. Rev. Lett. \textbf{92}, 145901 (2004).

\bibitem{india2} M. Agarwal, A. Ganguly, and Ch. Chakravarty
J. Phys. Chem. B, \textbf{113}, 15284 (2009).

\bibitem{ionicmelts} M. Agarwal and Ch. Chakravarty, Phys. Rev. E, \textbf{79}, 030202(R)
(2009).

\bibitem{buldwater} Zh. Yan, S. V. Buldyrev, and H. Eu. Stanley,
Phys. Rev. E, \textbf{78}, 051201 (2008).

\bibitem{chainfluids} T. Goel, Ch. N. Patra, T. Mukherjee, and Ch.
Chakravarty, J. Chem Phys. \textbf{129}, 164904 (2008).

\bibitem{klekelberg} W.P. Krekelberg, T. Kumar, J. Mittal, J.R.
Errington and T.M. Truskett, Phys. Rev. E \textbf{79} 031203
(2009).

\bibitem{klekelberg1} M. J. Pond, W. P. Krekelberg, V. K. Shen, J.
R. Errington and Th. M. Truskett, J. Chem. Phys. \textbf{131},
161101 (2009).

\bibitem{barbosapot} A. B. de Oliveira, P. A. Netz, T. Colla, and M. C. Barbosa, J. Chem.
Phys. \textbf{124}, 084505 (2006)

\bibitem{wejcp}  Yu. D. Fomin, N. V. Gribova, V. N. Ryzhov, S. M.
Stishov and Daan Frenkel, J. Chem. Phys., 129, 064512 (2008).

\bibitem{wepre} N. V. Gribova, Yu. D. Fomin, Daan Frenkel, V. N. Ryzhov,
Phys. Rev. E \textbf{79}, 051202 (2009).

\bibitem{gaussstill} F.H. Stillinger, J. Chem. Phys. \textbf{65},
3968 (1976).













\end{thebibliography}

\end{document}